\documentclass[twocolumn,english,prl,amssymb,aps,superscriptaddress,showpacs,twocolumn,amsmath,showkeys,floatfix]{revtex4-2}

\usepackage{geometry}
\usepackage[active]{srcltx}
\usepackage{textcomp}
\usepackage{amsmath}
\usepackage{graphicx}
\usepackage{color}
\usepackage{esint}

\makeatletter

\usepackage[english]{babel}
\usepackage{physics}
\usepackage[breaklinks]{hyperref}
\usepackage{inputenc}

\makeatother

\begin{document}

\title{Nonlinear dynamics of dissipative structures in coherently-driven Kerr cavities with a parabolic potential}

\author{Yifan Sun}\email{yifan.sun@uniroma1.it}
\affiliation{Department of Information Engineering, Electronics and Telecommunications, Sapienza University of Rome, Via Eudossiana 18, 00184 Rome, Italy}
\author{Pedro Parra-Rivas}
\affiliation{Department of Information Engineering, Electronics and Telecommunications, Sapienza University of Rome, Via Eudossiana 18, 00184 Rome, Italy}
\author{Mario Ferraro}
\affiliation{Department of Information Engineering, Electronics and Telecommunications, Sapienza University of Rome, Via Eudossiana 18, 00184 Rome, Italy}
\author{Fabio Mangini}
\affiliation{Department of Information Engineering, Electronics and Telecommunications, Sapienza University of Rome, Via Eudossiana 18, 00184 Rome, Italy}
\author{Stefan Wabnitz}
\affiliation{Department of Information Engineering, Electronics and Telecommunications, Sapienza University of Rome, Via Eudossiana 18, 00184 Rome, Italy}
\affiliation{CNR-INO, Istituto Nazionale di Ottica, Via Campi Flegrei 34, 80078 Pozzuoli, Italy}

\begin{abstract}
By means of a modified Lugiato-Lefever equation model, we investigate the nonlinear dynamics of dissipative wave structures in coherently-driven Kerr cavities with a parabolic potential. 
The potential stabilizes system dynamics, leading to the generation of robust dissipative solitons in the positive detuning regime, and of higher-order solitons in the negative detuning regime.
In order to understand the underlying mechanisms which are responsible for these high-order states, we decompose the field on the basis of linear eigenmodes of the system. This permits to investigate the resulting nonlinear mode coupling processes.
By increasing the external pumping, one observes the emergence of high-order breathers and chaoticons. 
Our modal content analysis reveals that breathers are dominated by modes of corresponding orders, while chaoticons exhibit proper chaotic dynamics.
We characterize the evolution of dissipative structures by using bifurcation diagrams, and confirm their stability by combining linear stability analysis results with numerical simulations.
Finally, we draw phase diagrams that summarize the complex dynamics landscape, obtained when varying the pump, the detuning, and the strength of the potential.

\end{abstract}
\maketitle


\section{Introduction}

Over the past decade, the generation and manipulation of dissipative temporal Kerr solitons (DKS) \cite{wabnitz:93} has become an increasingly important topic in photonics, particularly for their applications to fiber lasers and coherent frequency comb generation \cite{akhmediev_dissipative_2008,herr_temporal_2014-1,PASQUAZI20181}. Dissipative solitons differ from their conservative counterparts, as they require a balance between internal dissipation and external energy flow, in addition to the counter-balance between dispersion and nonlinearity. DKS dynamics and stability have been extensively analyzed within the mean-field approximation, whereby passive Kerr resonators are described by a temporal Lugiato–Lefever equation (LLE), which is a driven and damped, and detuned nonlinear Schroedinger model \cite{haelterman_dissipative_1992,PhysRevA.88.023819, chembo_theory_2017}. A variety of DKS can be formed in cavities operated in both the anomalous or in the normal dispersion regime \cite{parra-rivas_dark_2016,parra-rivas_bifurcation_2018-1,parra-rivas_origin_2021}. 
As the pump intensity increases, DKS undergo different types of instabilities, leading to complex spatiotemporal dynamics, which can be either periodic (i.e., breathers) or chaotic \cite{leo_temporal_2010,anderson_observations_2016,liu_characterization_2017,lucas_breathing_2017,bao_observation_2018}.

High-order effects, such as third-order dispersion, may considerably reduce the extension of unstable parameter regions, in favor of static DKS, and ultimately lead to the appearance of new types of localized states, as well as to the coexistence of bright and dark DKS \cite{tlidi_high-order_2010,parra-rivas_coexistence_2017,li_experimental_2020}. The Raman effect also plays a role in the dynamics of localized states \cite{parra-rivas_influence_2020}. 
Some modulation techniques may provide an additional degree of freedom for controlling and even suppressing instabilities.
Specifically, the spatiotemporal dynamics can be controlled by introducing an intensity or a phase modulation of the driving field  \cite{Wabnitz:96,Jang2015,Cole2018,Hendry2019,Talenti2020,Erkintalo2022}. Such inhomogeneous pumping enhances the pump-to-soliton conversion efficiency \cite{Erkintalo2022}, and provides additional deterministic routes for DKS generation, without undergoing an intermediate spatiotemporal chaotic phase \cite{Taheri2015}.
Another control method involves the direct modulation of the intracavity field, i.e., by adding an intracavity phase modulator \cite{Mecozzi:92,Tusnin2020,Englebert2021,yuan2021synthetic,Bersch2012}. In this case, one obtains a so-called \textit{synthetic dimension} \cite{yuan2021synthetic,Englebert2021}. 
Both methods may impress a phase modulation on the cavity field, thus creating an effective periodic potential for temporally confining optical pulses, for controlling the spatiotemporal cavity dynamics and its emerging dissipative states. Together with the stabilization of chaotic states \cite{Lobanov2016}, the potential may also lead to the emergence of \textit{chimera}-like states \cite{nielsen_engineered_2019,Tusnin2020,Sun2022}. Recently, we have discovered that these unstable dynamical states could be stabilized by adding a temporal parabolic potential in both 1D \cite{Sun2022} and 3D settings \cite{Sun2022b}.


In this work, we investigate the complex nonlinear dynamics arising in coherently-driven nonlinear Kerr cavities with synchronous phase modulation. 
We describe the latter through a parabolic potential in time. We find that the potential impacts on the dynamics of the system in an unexpected manner, leading to multiple robust high-order DKS that only exist for a particular parameter set.  
We investigate the origin of these higher-order DKS in the linear and nonlinear regimes of the system. In other words, the emergence of high-order DKS, which are of nonlinear nature, can be understood from a purely linear mode analysis. 
Besides, in the strong pump regime, we show the existence of different high-order breathers, which may be either symmetric or asymmetric in time, and we find that their dynamics can be elucidated from their linear mode components. 
Furthermore, we observe localized spatiotemporal chaotic states, also known as {\it chaoticons} or {\it chimera states} \cite{verschueren_chaoticon_2014}.
Finally, we summarize the observed complex dynamics by means of phase diagrams, which highlight the different types of system behavior.

This paper is organized as follows:
In Section \ref{sec_model}, we will introduce the mean-field model that we are going to use for describing our system. 
Section \ref{sec_scan} is dedicated to performing a phenomenological comparison of stable solitons under different conditions: specifically, in the absence or in the presence of a parabolic potential.
In Section \ref{sec_modal_analysis}, the origin of high-order DKS is analyzed by means of the linear eigenmodes of the cavity. In Section~\ref{DKS} we perform a bifurcation analysis of such states, by including nonlinearity. 
Section \ref{sec_bif_breathers} introduces breather solutions of different orders within a bifurcation diagram.
Section \ref{sec_chaoticon} demonstrates the chaotic nature of the chaoticons via the computation of the Lyapunov exponents of the modal energies.
Later, in Sections \ref{sec_pha} and \ref{sec_pot_str}, we unveil the organization of the different dynamical states in the form of phase diagrams. 
Finally, Section \ref{sec_conclu} presents the main conclusions of our paper.

\section{Model and normalization}\label{sec_model}

\begin{figure}[tb]
	\centering
	\includegraphics[width=\columnwidth]{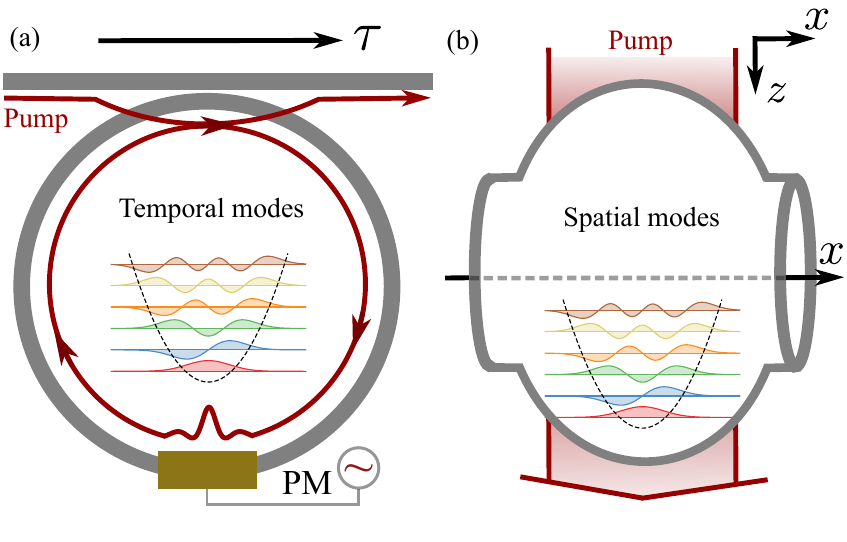}
	\caption{
		The physical model Eq.~(\ref{eq_main}) could be applied to a passive coherelty-driven cavity in (a) and to a bottle resonator in (b).
	}
	\label{fig_toy}
\end{figure}

In the mean-field approximation, the optical field envelope $\tilde{A}(\tau,t)$ of the electric field circulating within the ring resonator, in the presence of synchronous phase modulation [see Fig.~\ref{fig_toy}(a)] is governed by the modified Lugiato-Lefever equation (LLE) \cite{Englebert2021,Sun2022}:
\begin{equation}
	\begin{aligned} 
		\mathrm{i}t_{\rm R}\frac{\partial \tilde{A} }{\partial \tilde{t}} -& \frac{\beta_2L}{2}\frac{\partial ^2\tilde{A} }{\partial \tilde{\tau}^2}+\gamma L|\tilde{A}|^2\tilde{A}\\
		=&-\mathrm{i} \frac{\alpha}{2}\tilde{A} +\mathrm{i}\sqrt{\kappa P_{\mathrm{in}}}+ [\theta
		-V(\tilde{\tau})]\tilde{A}\\
	\end{aligned}
	\label{eq_with_units}
\end{equation}
where $t_{\rm R}$ is the round-trip time of pulses in the resonator, $\tilde{\tau}$ is a fast time, which describes the temporal profile of the intracavity waveform, in a reference frame moving with the group velocity of light, $\tilde{t}$ is a slow time, a continuous version of the time measuring intervals between successive round trips in the cavity, $\beta_2$ is second-order or chromatic dispersion, $L$ is the cavity length, $\gamma$ is the nonlinear coefficient per round-trip, $\alpha$ is the loss per round-trip, $P_{\rm in}$ input power of the homogeneous pumping, $\kappa$ is the coupling coefficient from the pump into the cavity, $\theta$ is s  the mean-value of the phase detuning between the driving field and a cavity resonance, and $V(\tilde{\tau})$ represents the action of synchronous phase modulation. 

The presence of a synchronous phase modulation has been generally modeled by introducing a cosine-type potential of the form $V(\tilde{\tau}) = J_{\rm M}\cos(\Omega_{\rm M} \tilde{\tau})$, where $\Omega_{\rm M}$ and $J_{\rm M}$ are the modulation frequency and the depth of the phase modulator, respectively. \cite{Tusnin2020,Englebert2021}. This potential can be Taylor-expanded as $V(\tilde{\tau})=J_{\rm M}(1-\frac{1}{2}\Omega_{\rm M}^2\tilde{\tau}^2+\frac{1}{24}\Omega_{\rm M}^4\tilde{\tau}^4)+\dots$. In this work, we focus on the 
limited regime $\Omega_{\rm M}\tilde{\tau}\ll1$, where the potential is approximately expressed as 
\begin{gather}
    V(\tilde{\tau}) = J_{\rm M}(1-\frac{1}{2}\Omega_{\rm M}^2\tilde{\tau}^2).
\end{gather}



In this regime, Eq.~(\ref{eq_with_units}) can be rewritten by using the normalized scale transformations and dimensionless parameters  
\begin{equation}
	\begin{aligned} 
		t 		&= \tilde{t}\alpha/2t_R \\
		\tau 	&= \tilde{\tau}  \sqrt{\alpha/|\beta_2|L} \\
		A 		&= \tilde{A}\sqrt{2\gamma L/\alpha} \\
		P 		&=  \sqrt{8\kappa P_{\rm in}\gamma L/\alpha^3} \\
		\delta 	&= 2(\theta-J_{\rm M})/\alpha \\
		C		&= J_{\rm M}\Omega_{\rm M}^2|\beta_2| L/\alpha^2. \\		
	\end{aligned} 
\end{equation}
In this way, we may transform Eq.~(\ref{eq_with_units}) into the dimensionless equation
\begin{gather} 
	\frac{\partial A}{\partial t} = \mathrm{i}\frac{\partial^2}{\partial \tau^2}A - \mathrm{i}C\tau^2A 
	+ \mathrm{i}|A|^2A  - (1+\mathrm{i}\delta)A+ P,
	\label{eq_main}
\end{gather}
where three dimensionless variables govern the dynamics of the cavity: the cavity detuning  $\delta$, the pump $P$, and the potential strength $C$.

It is worth noting that the same model (Eq.~(\ref{eq_main}) can be applied to describe bottle microresonators [see Fig.~\ref{fig_toy}(b)] \cite{Oreshnikov2017,Kartashov2018,Sumetsky2019}, where the potential term results from the bump shape of the bottle structure or the inhomogenous pump beam. In this case, one needs to replace the fast time variable $\tau$ with the spatial variable $x$.
In both cases, parabolic potentials create Hermite-Gaussian modes in the temporal or spatial domain [see Fig.~\ref{fig_toy}].
Besides, parabolic potentials have been previously considered in conservative systems for studying, e.g., vortex solitons in Bose-Einstein condensate \cite{malomed_stability_2007,lashkin_stable_2008}, and in dissipative systems for studying mode-locked nanolasers\cite{Sun2019, Sun2020}, multimode fiber lasers \cite{Kalashnikov_2022}, and for stabilizing 3D solitons \cite{Sun2022b}.

\begin{figure*}[!t]
	\centering
	\includegraphics[width=\textwidth]{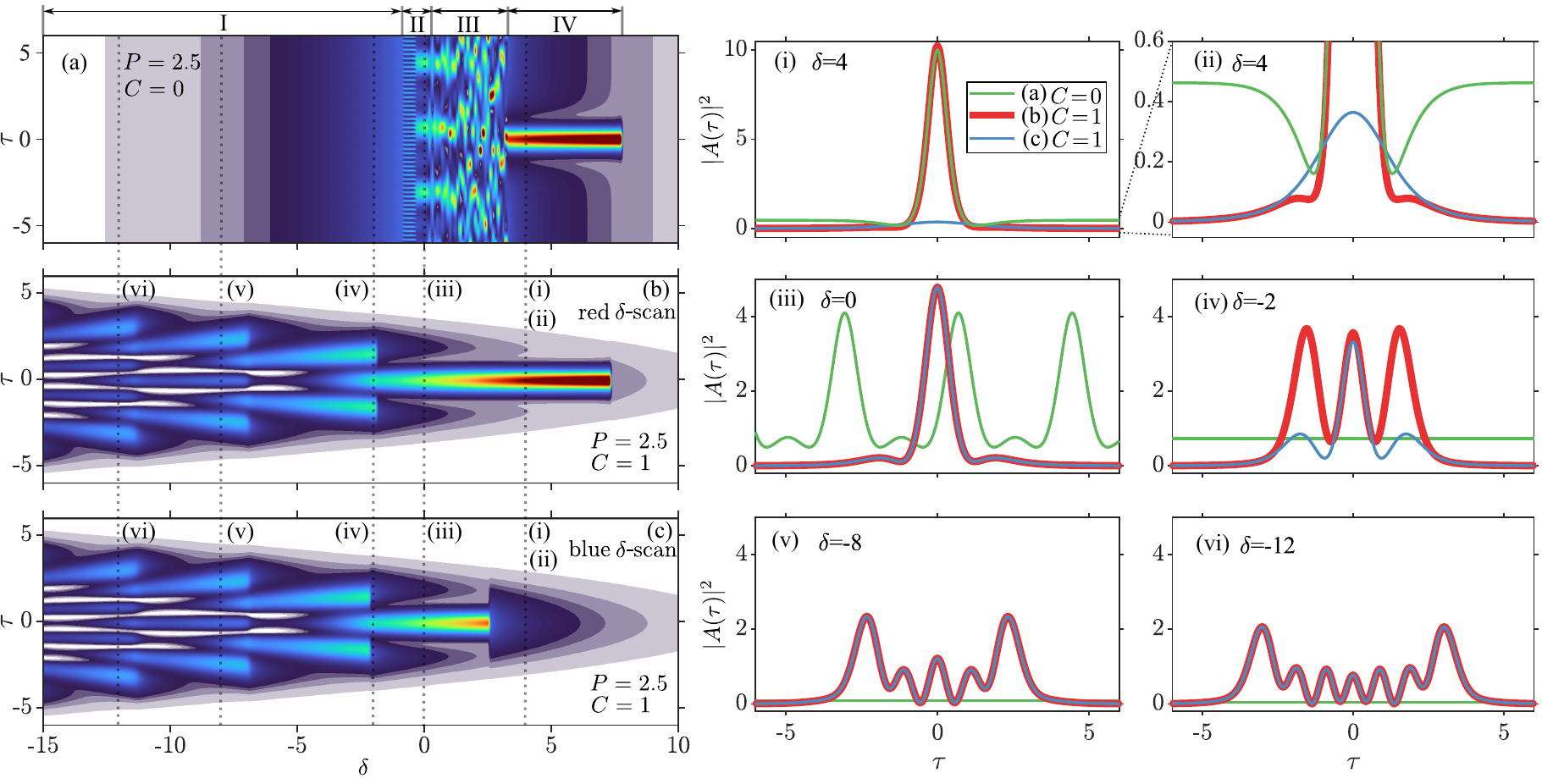}
	\caption{
		In the left panels, we compare the field power profile $|A(\tau)|^2$ in the absence of the potential ($C=0$, red $\delta$-scan) in (a), or in the presence of the potential  ($C=1$, red $\delta$-scan) in (b) and ($C=1$, blue $\delta$-scan) in (c), when varying the detuning $\delta$.
		In the center and right panels, we trace the field intensity  $|A(\tau)|^2$, with or without the potential, for selected values of the cavity detuning $\delta$: green curves refer to $C=0$, red curves to $C=1$, red $\delta$-scan, and blue curves to $C=1$, blue $\delta$-scan; $\delta=4$ in (i,ii), $\delta=0$ in (iii), $\delta=-2$ in (iv), $\delta=-8$ in (v), $\delta=-12$ in (vi).
	}
	\label{fig_pump2p5}
\end{figure*}
To compute the solutions of Eq.~(\ref{eq_main}), direct numerical simulations by a pseudo-spectral method, and numerical continuation through AUTO-07p\cite{Doedel2009}. Both methods generate the same results.
For the latter, a linear stability analysis will also be used, in order to evaluate the solution stability. 
In order to do continuation of Eq.~(\ref{eq_main}) by using AUTO-07p \cite{Doedel2009}, the model of Eq.~(\ref{eq_main}) needs to be represented in the format of an ordinary differential equation (ODE), as we have described in Appendix~\ref{AppB}.


\section{Influence of the potential on dissipative soliton dynamics}\label{sec_scan}


Let us start by considering the impact of the potential on cavity dynamics. This is done by comparing different solutions, with or without the potential. 
The colormap in Fig.~\ref{fig_pump2p5}(a) shows the steady-state field power of numerical solutions of Eq.~(\ref{eq_main}) without the potential term $-\mathrm{i}C\tau^2A$ (setting $C=0$), when the cavity detuning $\delta$ gradually increases, and $P=2.5$. Here we call it a red $\delta$-scan, since it corresponds to increasing the wavelength (red-shift) of the pump. 
This red $\delta$-scan process is carried out in the following manner: we use a final steady-state solution (a solution until convergence is obtained) $A(\tau,t=t_{\rm final})$ with a detuning $\delta_n$ as the initial condition $A(\tau,t=0)$ with the new detuning $\delta_{n+1}$, where $\delta_{n+1}>\delta_{n}$. In this way, one may obtain all the steady-state solutions (instead of transient solutions), with a significant reduction of the convergence time in simulations.
Fig.~\ref{fig_pump2p5}(a) shows that the intra-cavity field starts from the homogeneous solution at $\delta=-15$: when the detuning increases, $|A(\tau)|^2$ increases from $0.03$ up to $1.17$ at $\delta=-0.9$ in the region I [see the marks on the top of Fig.~\ref{fig_pump2p5}(a)]. 
By keeping increasing $\delta$, Turing patterns occur in region II between $-0.9\le\delta\le0.2$, then chaotic patterns appear in region III between  $0.2<\delta<3.2$. The occurrence of stable, time-localized dissipative soliton solutions is observed in region IV, between $3.3\le\delta\le7.7$.


However, when the parabolic potential $-\mathrm{i}C\tau^2A$ is included, different dynamics are discovered. Fig.~\ref{fig_pump2p5}(b) shows a red $\delta$-scan, analogous to the one depicted in Fig.~\ref{fig_pump2p5}(a), but now for $C=1$. As we can observe, the presence of the potential leads to a clear difference in the intra-cavity dynamics. 
To better compare the cases without and with the potential, in Fig.~\ref{fig_pump2p5}(i-vi) we plot the power distribution $|A(\tau)|^2$ of the different states for the same parameter set, with the potential (see red curves) and without the potential (see green curves).
For the stable soliton regime, the profiles are very similar, except that the homogeneous background is suppressed by the potential (see Fig.~\ref{fig_pump2p5}(i,ii)).
In addition, the parabolic potential also stabilizes the unstable regimes of region II and III in Fig.~\ref{fig_pump2p5}(a), leading to the formation of a stable soliton (see Fig.~\ref{fig_pump2p5}(iii), for $\delta=0$). We have reported this stabilization in a previous work \cite{Sun2022}. 

More interestingly, the potential also extends the stable DKS existence into a large negative detuning region [see Fig.~\ref{fig_pump2p5}(b)]: without the potential, only homogeneous steady-states exist (cfr. with  Fig.~\ref{fig_pump2p5}(a)). Figures Fig.~\ref{fig_pump2p5}(iv)-(vi) compare these two scenarios (plotted by the green curves for $C=0$ and the red curves for $C=1$) for $\delta=-2,\,-8,\,-12$, respectively. 
The difference in DKS order manifests in the number of side peaks which are present in each state: as we can see, the number of peaks changes when increasing $\delta$ in Fig.~\ref{fig_pump2p5}(b). Note that, when decreasing $\delta$, 
the power of the DKS side peaks varies significantly with their order.

\begin{figure}[t]
	\centering
	\includegraphics[scale=1]{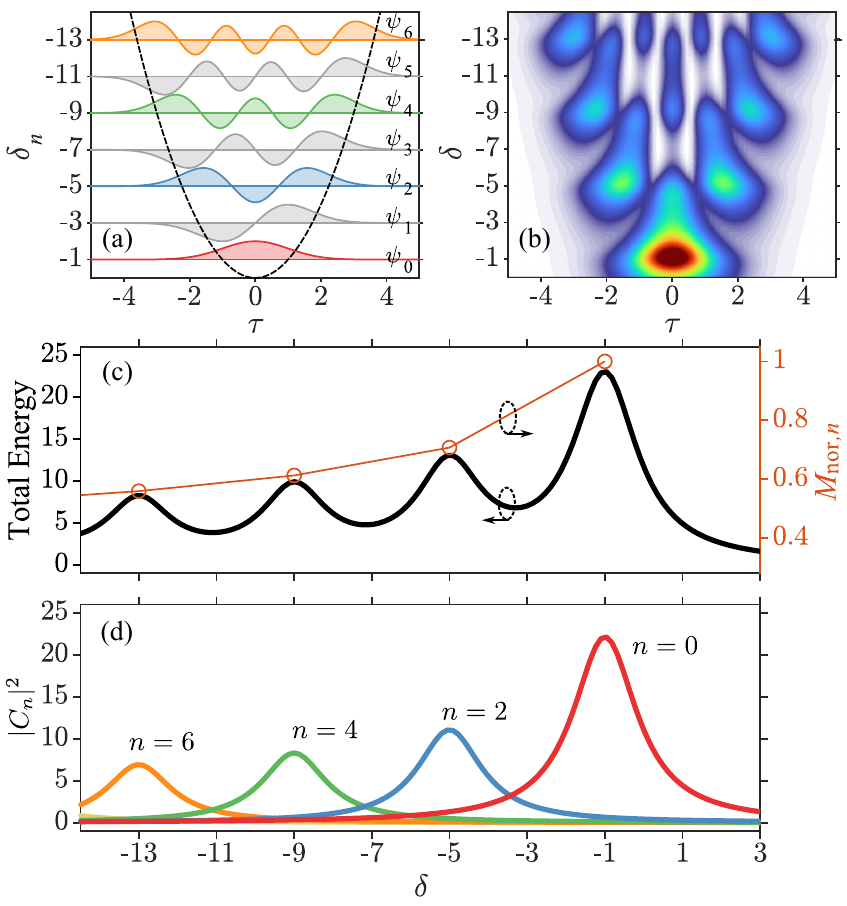}
		\caption{
			(a) Hermite-Gauss eigenmodes $\psi_n(\tau)$ (solid curves) located in correspondence of their eigenvalues $\delta_n$, and the parabolic potential $-C\tau^2$ (dashed black curves) of the system, when $C=1$. 
			(b) Linear stable solutions $|A(\tau)|^2$ in the absence of the Kerr effect, vs. detuning $\delta$. 
			(c) Even mode excitation efficiency $M_{\mathrm{nor},2n}$ (plotted by circles) and total energy of solutions as in panel (b), vs. detuning  $\delta$. 
			(d) Modal energies $|C_n|^2$ vs. detuning $\delta,$ by decomposing the solutions in (b) on the eigenmodes in (a).}
	\label{fig_mode_analysis}		
\end{figure}

We have also implemented a $\delta$-blue scan (by progressively reducing the detuning $\delta$). The result is illustrated in Fig.~\ref{fig_pump2p5}(c). 
The individual examples with corresponding $\delta$ are plotted as blue curves in Fig.~\ref{fig_pump2p5}(i-vi), respectively.
For $\delta>2.6$, a small amplitude localized state, which does not exist in the $\delta$-red scan, appears [see, e.g., the one depicted in blue in Fig.~\ref{fig_pump2p5}(ii)]. This state corresponds to a modification of the homogeneous background state.
For $\delta<2.6$, the DKS appearance follows the same sequence as in Fig.~Fig.~\ref{fig_pump2p5}(b) [see for example the states plotted by the red and blue curves in Fig.~\ref{fig_pump2p5}(iii) for $\delta=0$]. When decreasing $\delta$ even further, DKS of different order may coexist for the same range of $\delta$ values (e.g., for $-2.1<\delta<-1.8$)
as we can appreciate from the discrepancy between Fig.~\ref{fig_pump2p5}(b) and Fig.~\ref{fig_pump2p5}(c) [see in more detail Fig.~\ref{fig_pump2p5}(iv) for $\delta=-2$]. All of these states are quite robust and strong attractors of the system.

At this stage, we may ask ourselves a few questions, such as: how are these different DKS organized from a bifurcation perspective, and how does their nature change with the control parameters of the system? Our main aim in the rest of this paper is to answer, step by step, these two questions.

\section{Modal analysis in the absence of Kerr nonlinearity}\label{sec_modal_analysis}



In this section, we will analyze the linear eigenmodes of the cavity field, which give rise to the high-order resonant states, enabling us to uncover some of the essential features of the system. Subsequently, we will derive the coupled mode equations of the model from Eq.~(\ref{eq_main}) to demonstrate the different pumping rates for each mode. Finally, we will examine the linear resonance of the cavity by ignoring the Kerr effect.


Let us first ignore, besides nonlinearity, also the effects of pumping, dissipation, and cavity detuning. This leads to the linear eigenvalue equation  
\begin{equation} 
	\hat{H_0}\psi_n(\tau) = \mathrm{i}\delta_n\psi_n(\tau),
 \label{eq_eigen_fu_va}
\end{equation}
where  $\hat{H_0}=\mathrm{i} \partial_\tau^2  - \mathrm{i}C\tau^2$ is a linear operator, accounting for the presence of second-order or chromatic dispersion and the parabolic potential: 
$\psi_n$ and $\delta_n$ are the corresponding eigenmodes and eigenvalues. The latter, determining the resonance frequencies, are equally spaced, given by
\begin{equation}
	\delta_n=-2\sqrt{C}(n+1/2).
	\label{eq_eigenvalue}
\end{equation}
The former are Hermite-Gauss (HG) functions, given by 
\begin{equation} 
	\psi_n(\tau) = (2^nn!)^{-\frac{1}{2}}\pi^{-\frac{1}{4}}\exp\left(-\frac{\tau^2}{2a_\tau^2}\right) H_n\left(\frac{\tau}{a_\tau}\right)
\end{equation}
where the scaling factor is
\begin{equation}
	a_\tau=C^{-1/4},
	\label{eq_scaling}
\end{equation}
and $H_n$ is the Hermite polynomial. The relation between eigenfunctions $\psi_n(\tau)$, eigenvalues $\delta_n$ and the parabolic potential $-iC\tau^2$ is illustrated in Fig.~\ref{fig_mode_analysis}(a). 


Eigenmodes $\psi_n(\tau)$ satisfy the orthonormality condition $$\int\psi_n(\tau)\psi_m(\tau)\dd \tau=\delta_{nm},$$ and since they form a complete orthogonal basis, any time evolution solution $A(\tau,t)$ of Eq.~(\ref{eq_main}) can be expanded on the HG mode basis,
\begin{equation} 
	A(\tau,t) = \sum_{n=0}^NC_n(t)e^{i\delta_nt}{ \psi}_n(\tau),
	\label{A_ex}
\end{equation}
where the mode complex amplitude reads
\begin{equation} 
	C_n(t)e^{i\delta_nt} = \int_{-\infty}^{\infty}A(\tau,t)\psi_n(\tau)\dd \tau,
	\label{eq_cn}
\end{equation}
and the mode energy corresponds to $|C_n|^2$.

To continue further with our modal analysis, let us focus on how the pump influences the evolution of each mode separately. 
To do so, we first substitute Eq.~(\ref{A_ex}) into Eq.~(\ref{eq_main}) and project on the mode $\psi_n$. This leads [see Appendix~\ref{App_A}] to a set of ODEs 
\begin{equation} 
	\begin{split}
		\frac{\dd C_n(t)}{\dd t} =	&e^{-i\delta_nt}\int_{-\infty}^{\infty}P\psi_n(\tau)\dd \tau\\
		+&e^{-i\delta_nt}\int_{-\infty}^{\infty}\mathrm{i}|A|^2A\psi_n(\tau)\dd \tau\\
		-&e^{-i\delta_nt}(1+\mathrm{i}\delta)C_n.\\
	\end{split}
\label{eq_couple_mode}
\end{equation}
Note that these ODEs govern the mode dynamics, which is equivalent to the field dynamics in Eq.~(\ref{eq_main}). Now each ODE governs the dynamics of one mode, while the interaction of modes results from the nonlinear coupling term (the second term in Eq.~(\ref{eq_couple_mode})).

The first term in Eq.~(\ref{eq_couple_mode}) corresponds to the pumping rate into each mode. 
Ignoring the factor leading to a phase change, this pumping rate
\begin{equation} 
	M_n(\tau) = \int_{-\infty}^{\infty} P\psi_n(\tau) \dd \tau,
\end{equation}
depends on the mode amplitude distribution [see Fig.~\ref{fig_mode_analysis}(a)]. From here, we can define the {\it mode excitation efficiency} via the pumping rate normalization $M_{\mathrm{nor},n}=M_n/M_1$.

The $\tau$-asymmetry of odd modes ($n=1,\,3,\,5,\,\cdots$) implies that $M_{\mathrm{nor},2n+1} = 0$, since opposite amplitude components with respect to $\tau=0$ cancel their contribution to the mode excitation efficiency.
Therefore, odd modes are suppressed by the coherent pump. For even-symmetric modes ($n=0,\,2,\,4,\,\cdots$), in contrast, the exciting ratios $M_{\mathrm{nor},2n}$ are non-vanishing and they decrease with the eigenvalue $\delta_{2n}$ (i.e., increasing $n$), as  depicted by the circles in Fig.~\ref{fig_mode_analysis}(c). 

These differences in mode excitation ratios can be confirmed by performing numerical simulations of Eq.~(\ref{eq_main}), without the nonlinear Kerr term $\mathrm{i}|A|^2A$.
In order to do so, we compute the steady-state solutions of the system as a function of $\delta$ for $P=2.5$. The field power distribution  $|A(\tau)|^2$ resulting from these computations is depicted in Fig.~\ref{fig_mode_analysis}(b).  
The shape of this distribution is associated with the mode shapes of 
Fig.~\ref{fig_mode_analysis}(a). As a result, the field power distribution expands its temporal range as $\delta$ decreases, and it is symmetric with respect to $\tau=0$. 
The similarity between Fig.~\ref{fig_mode_analysis}(a) and Fig.~\ref{fig_mode_analysis}(b) indicates that the resonant contribution predominantly comes from even modes, while odd modes are suppressed.

This is more obvious when we plot the corresponding total field energy $E(\delta) = \iint|A(\tau,\delta)|^2\dd\tau$ [see the black curve in Fig.~\ref{fig_mode_analysis}(c)]
Here, the resonance peaks only appear at $\delta$ values corresponding to even mode eigenvalues. When compared with the excitation efficiency $M_{\mathrm{nor},2n}$, these peaks exhibit a very similar tendency.
Furthermore, we can decompose all of the linear states on the HG mode basis by using Eq.~(\ref{eq_cn}). The results are depicted in Fig.~\ref{fig_mode_analysis}(d), where the mode energies $|C_n|^2$ are traced vs. detuning $\delta$. From here, we can conclude that the linear cavity resonances are related to the presence of even HG modes only, since all odd modes are left unexcited by the pump.


%

\section{Bifurcation structure of high-order dissipative Kerr solitons}\label{DKS}
\begin{figure}[!t]
	\centering	\includegraphics[width=\columnwidth]{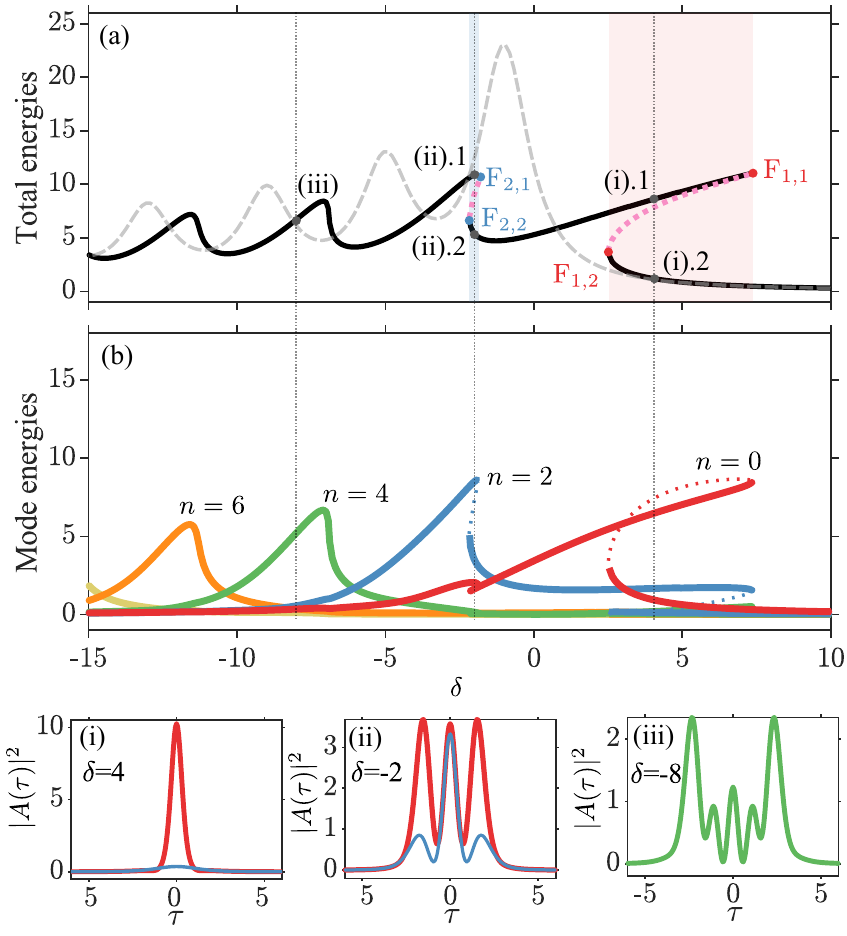}	
	\caption{Bifurcation diagram vs. $\delta$, showing the total energy of stable (unstable) solutions by solid black (pink dot) curves in (a), and their mode energies $|C_n|^2$ in (b) for the parameters $(P,C) = (2.5,1)$. For comparison, linear solutions in the absence of the Kerr effect are also plotted by the dashed gray curve in (a). The regime of bistable solutions, created by fold bifurcations (F), are marked by the transparent colored regions.
	}
	\label{fig_bif_pump_2p5}		
\end{figure}

At this point, we are ready to introduce the Kerr nonlinearity $\mathrm{i}|A|^2A$, which permits us to analyze the emergence of DKS, and nonlinear modifications of cavity resonances. Kerr nonlinearity leads to an intensity-dependent nonlinear phase shift, so that resonance peaks shift towards positive $\delta$ (i.e., $\delta\rightarrow \delta +|A|^2$). Therefore, the larger $|A|^2$, the more prominent the shift. An example of this nonlinear resonance is illustrated in Fig.~\ref{fig_bif_pump_2p5}(a) for $P=2.5$. To compute this bifurcation diagram, we need to follow a path-parameter continuation approach, which allows us to numerically compute the nonlinear steady-state solutions, such as DKS, of the cavity field. Indeed, each point on the diagram of Fig.~\ref{fig_bif_pump_2p5}(a) corresponds to a different DKS state. The linear stability of these states is shown by using solid (dashed) lines for stable (unstable) states. 


For a better comparison, in Fig.~\ref{fig_bif_pump_2p5}(a) we also plot the linear resonances by using a dashed gray curve. The DKS modify along this bifurcation diagram, as depicted in Fig.~\ref{fig_bif_pump_2p5}(i)-(iii).
The resonance peaks associated with the DKS not only shift, but also tilt towards positive values of $\delta$. Such tilts may lead to the appearance of bistable regions, where two stable states coexist [see the transparent red and blue regions in  Fig.~\ref{fig_bif_pump_2p5}(a)]. This is the case of the small and high-amplitude localized states shown in blue and red, respectively, in Fig.~\ref{fig_bif_pump_2p5}(i). This bistable region extends between the fold (F) bifurcations $\rm F_{1,1}$ and $\rm F_{1,2}$, which are connected through an unstable DKS branch [see the dashed pink line in Fig.~\ref{fig_bif_pump_2p5}(a)]. When decreasing $\delta$, the high-amplitude DKS become unstable at F$_{2,2}$ and start to nucleate two side peaks. Following up the resonance, the side peaks grow larger in amplitude, until eventually, the DKS stabilizes at F$_{2,1}$. The DKS at this stage looks like the one which is plotted in Fig.~\ref{fig_bif_pump_2p5}(ii). The $\delta$ interval between these points corresponds to a second bistable DKS region. 
By decreasing $\delta$ even further, the DKS morphology becomes more complex, since new subsidiary peaks appear. An example of such states is shown in Fig.~\ref{fig_bif_pump_2p5}(iii).  

At this point, we may make use of what we learned in Section~\ref{sec_modal_analysis}, and analyze the DKS by projecting them on the HG mode basis, all along the nonlinear resonances of Fig.~\ref{fig_bif_pump_2p5}(a). By doing that, we may compute the modification of the mode energy $|C_n|^2$ as a function of $\delta$, for different mode orders ($n$) [see Fig.~\ref{fig_bif_pump_2p5}(b)]. This leads to observing that all DKS have multimode components, although for every resonance peak, there is always a single mode that dominates. Moreover, for two stable solutions between any two consecutive fold bifurcations, there is a sharp change of modal energy with respect to the corresponding resonance modes. For example, the two solutions (ii).1 and (ii).2 in Fig.~\ref{fig_bif_pump_2p5}(a,ii), located on the second resonance peak where the mode $n=2$ is dominant, exhibit different modal contents. The same holds for the first resonance peak.

As we will see in Section~\ref{sec_bif_breathers} and \ref{sec_pha}, by increasing the pump amplitude $P$, one increases the overall field power, until new bistable regions appear, leading to more complex spatiotemporal dynamics.

\section{Breathers of different orders}\label{sec_bif_breathers}

In the standard LLE model (i.e., Eq.(4) with $C=0$), increasing $P$ gives rise to dynamical instabilities, which lead to the formation of time-evolving breathers and chaos \cite{leo_dynamics_2013,parra-rivas_dynamics_2014}. In this section, we will explore how the potential modifies such dynamics, as well as the appearance of high-order breathers.

\begin{figure}[!t]
	\centering
\includegraphics[width=\columnwidth]{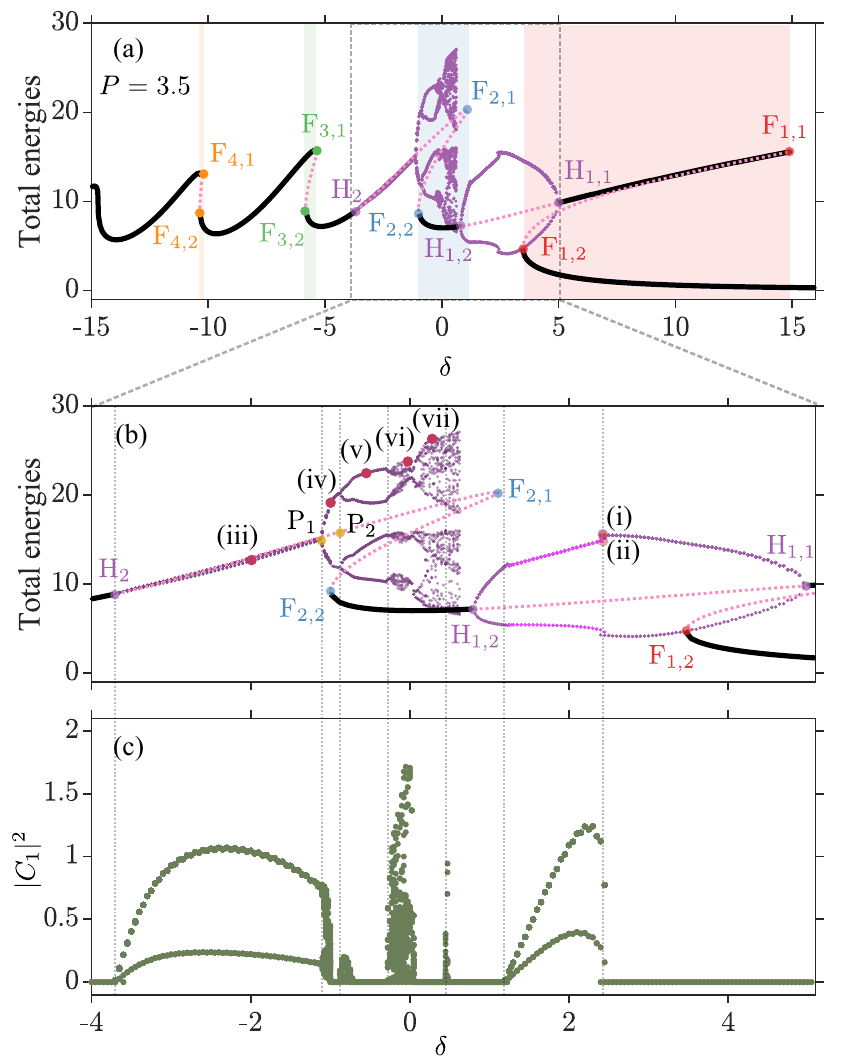}	
	\caption{
		Bifurcation diagram vs. $\delta$, showing the total energy of stable (unstable) solitons by solid black (pink dot) curves in (a), and the mode energy $|C_1|^2$ in (c) for the parameters $(P, C) = (3.5,1)$. A zoomed region between $-4\le\delta\le5.1$ in (a) is shown in (b). Fold and Hopf bifurcation points are marked at corresponding positions by F and H. The individual breather solutions at different $\delta$ are marked by the points (i-vii), and are plotted in Fig.~\ref{fig_breathers}.
		}	
	\label{fig_bif_3p5}
\end{figure}

Figure~\ref{fig_bif_3p5}(a) shows the bifurcation diagram of DKS for $P=3.5$. Although it is similar to that shown in Fig.~\ref{fig_bif_3p5}(a), now the resonance peaks tilt even for lower values of $\delta$, leading to the formation of fold bifurcations $\rm F_3$, $\rm F_4$ at the third and fourth resonance peaks. For the first two resonance peaks, the linear stability analysis, in this case, shows that DKS states undergo several Hopf bifurcations ($\rm H_{1,1}$, $\rm H_{1,2}$, $\rm H_{2}$), which leads to the emergence of breathing dynamics. 

\begin{figure*}[tpb]
	\centering
 \includegraphics[width=\textwidth]{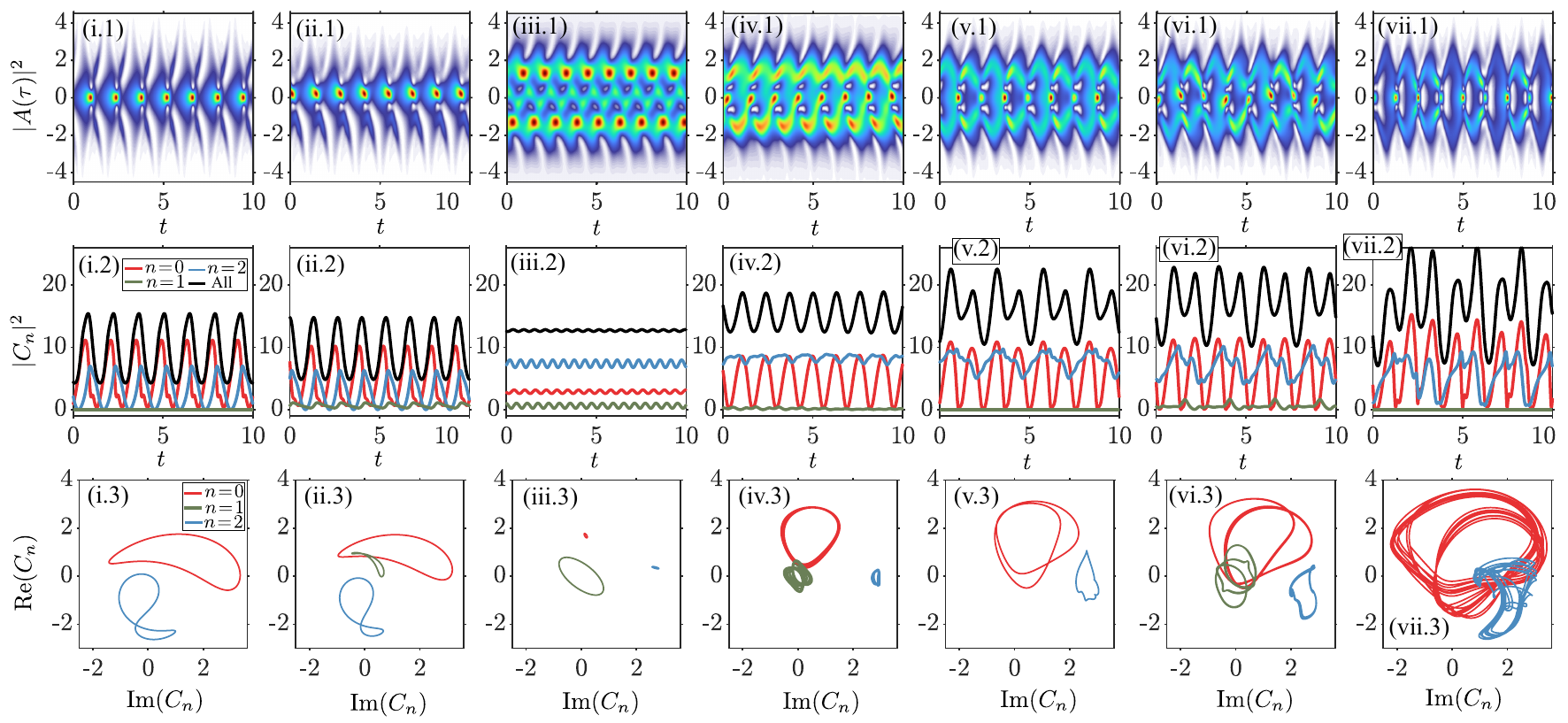}	
	\caption{Different breathers, corresponding the bifurcation diagram in Fig.~\ref{fig_bif_3p5}, are plotted here for (i,ii) $\delta=2.4$, (iii) $\delta=-2$, (iv) $\delta=-1$, (v) $\delta=-0.5$, (vi) $\delta=0$, (vii) $\delta=0.5$, when $P=3.5$, $C=1$. Sub-figures on the top, middle, and bottom panels, marked by (1), (2) and (3), respectively, correspond to the time evolution of power $|A(\tau,t)|^2$, total energy $\iint|A(\tau,t)|^2\dd \tau$ and modal energies $|C_n(t)|^2$, and to the time evolution of real and imaginary parts of the mode amplitudes $C_n(t)$, for each $\delta$ in (i-vii). 
	}
	\label{fig_breathers}
\end{figure*}

Let us take a closer look at this dynamical regime [see the close-up view of the region $-4\le\delta\le5$ shown in Fig.~\ref{fig_bif_3p5}(b)]. When decreasing $\delta$, the first instability that appears is $\rm H_{1,1}$. Here stable single-peak DKS destabilize, leading to breather states such as that of Fig.~\ref{fig_breathers}(i.1), where we show the evolution of $A(\tau)$ with $t$ for $\delta=-2.4$. In order to represent the modification of the breathers with $\delta$, we plot (by using black dots) the maxima and minima of the breather energy evolution vs. time $t$: $E(t)=\int|A(\tau,t)|^2\dd \tau$. This also allows us to distinguish among different breathing behaviors. 

The power of the breather at $\tau=0$, i.e., $|A(\tau=0,t)|^2$, exhibits periodic oscillations, as it can be appreciated in Fig.~\ref{fig_breathers}(i.1). Fig.~\ref{fig_breathers}(i.2) shows the periodic temporal evolution of the breather energy $E$, and the modal energies of the first three modes (i.e., $|C_{0,1,2}|^2$ ) by using black, red, dark green and blue curves, respectively. These periodic oscillations correspond to the nearby trajectories which are depicted in the reduced phase space plane \{$\mathrm{Re}[C_n(t)], \mathrm{Im}[C_n(t)]\}$ that is shown in Fig.~\ref{fig_breathers}(i.3).




By reducing $\delta$ even further [see Fig.~\ref{fig_bif_3p5}(b)], $\tau$-asymmetric breathers appear within the interval $\delta$-interval $1.23\le\delta\le2.4$: e.g., see Fig.~\ref{fig_breathers}(ii.2). At the frontier between these symmetric and asymmetric states, we find a narrow region where they coexist [see the close-up view in Fig.~\ref{fig_bif_3p5}(b)]. While the evolution of modes $\psi_{0,2}$ is similar for both symmetric and asymmetric states, the contribution of the odd mode $\psi_1$ only appears in the latter [see the purple curve in Fig.~\ref{fig_breathers}(ii.2) and (b.3)]. Therefore, asymmetric breathers can be easily identified by tracing out the odd modal energy components. To better clarify this point, in Fig.~\ref{fig_bif_3p5}(c) we plot the modification of $|C_1(t)|^2$ with $\delta$ for all oscillatory states. Thus, when $|C_1(t)|^2=0$ the DKS state is symmetric, and asymmetric otherwise. One clearly notices that, by reducing $\delta$, the asymmetric breathers become symmetric again at $\delta=1.23$. This last breather dies out at $\rm H_{1,2}$ ($\delta=0.78$), and then a single-peak DKS persist until $\rm F_{2,2}$ ($\delta=-1$) [see Fig.~\ref{fig_bif_3p5}(b)].

Asymmetric breathers emerge due to a balance between the mode suppression of homogeneous pumping and asymmetric mode excitation by Kerr nonlinearity. 
Therefore, these breathers usually appear when the pumping is relatively strong, and $\delta$ approaches a regime between two adjacent symmetric modes, where the asymmetric modes are favored. 

The breathers arising from H$_2$ undergo a Feigenbaum-cascade type of diagram, as shown in purple in Fig.~\ref{fig_bif_3p5}(a),(b), where a period-doubling cascade eventually leads to temporal chaos \cite{parra-rivas_dynamics_2014,Ferre2017}. In these states, the energy component that is associated with $\psi_2$ is well populated, as shown in Fig.~\ref{fig_breathers}(iii) for $\delta=-2$.
Owing to the contribution of asymmetric modes, the breather exhibits zigzag-type-like oscillations, where most of the energy on one side is exchanged with the other side [see Fig.~\ref{fig_breathers}(iii.1)]. 
For these breathers, the energy exchange between modes is much larger than its total energy fluctuation [see Fig.~\ref{fig_breathers}(iii.2)]. 
Therefore, in the region between $\rm H_2$ and $\rm P_1$, fluctuations of the total energy are very small [see Fig.~\ref{fig_bif_3p5}(b)], while the modal energy fluctuations (e.g., associated with $\psi_1$) are relatively large [see Fig.~\ref{fig_bif_3p5}(c)].
For $\delta< \delta_{\rm H_{2}}$, $\psi_2$ is dominant, and DKS are similar to that which is depicted in red in Fig.~\ref{fig_bif_pump_2p5}(ii).

By further increasing $\delta$, a period-doubling bifurcation (P$_2$) occurs at $\delta=-0.85$ [see Fig.~\ref{fig_bif_3p5}(b)], and the breather diagram splits into four branches. Two examples of 
symmetric and asymmetric breathers in this regime are shown in Fig.~\ref{fig_breathers}(v,vi) for $\delta=-0.5,\,0$, respectively. The period-doubling process continues and eventually, states evolve into the chaos-like breather states shown in Fig.~\ref{fig_breathers}(vii) for $\delta=0.5$.  
Between $\rm F_{2,2}$ and $\rm H_{1,2}$ [see Fig.~\ref{fig_bif_3p5}(b)], breather states coexist with DKS, which are eventually met by continuously increasing $\delta$.



\section{Chaoticons}\label{sec_chaoticon}

\begin{figure}[!t]
	\centering
	\includegraphics[width=\columnwidth]{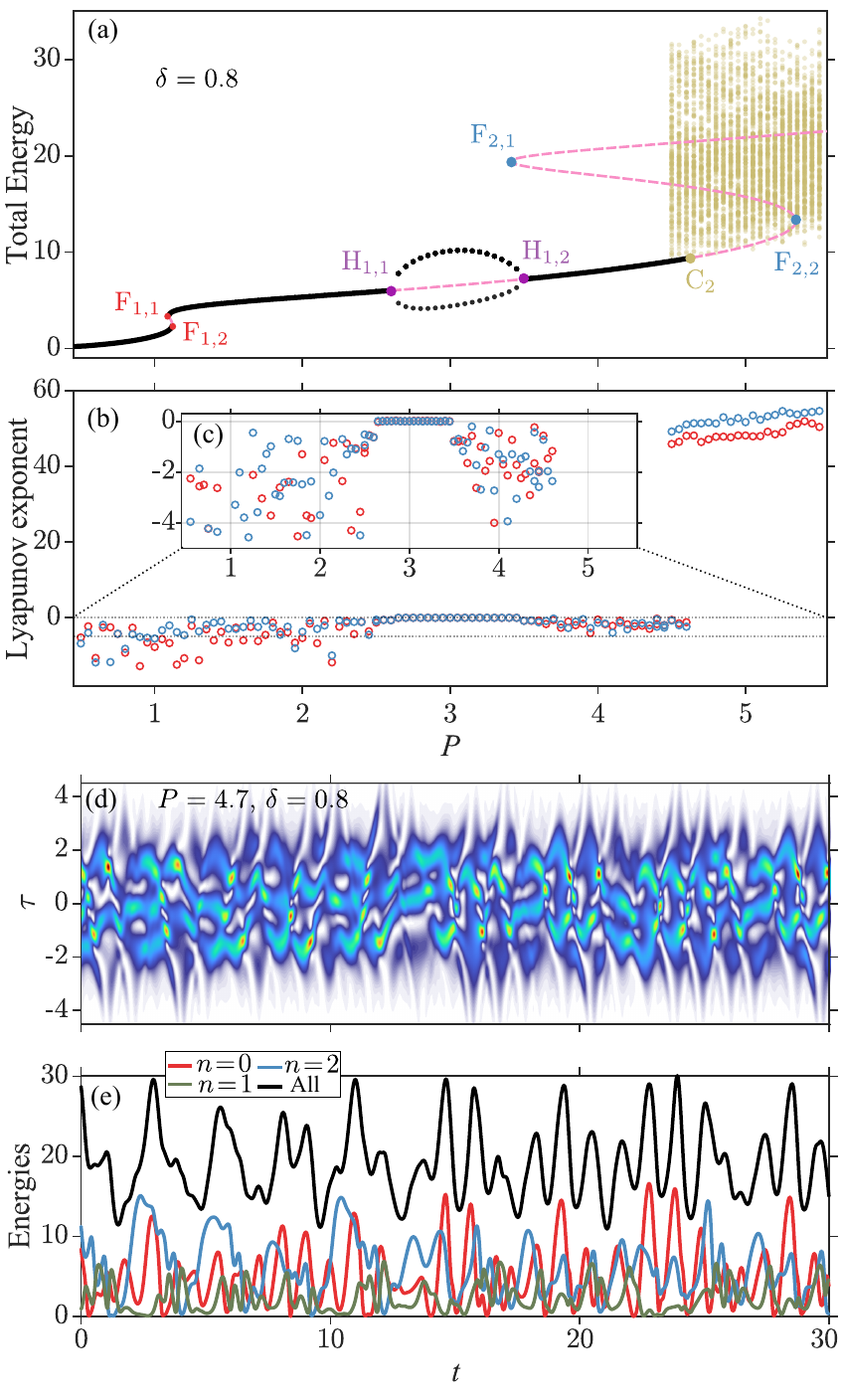}		
	\caption{Bifurcation diagram, showing the total field energy vs. $P$ in (a), and the largest Lyapunov exponent of modal energy associated with mode 0 (red circles) and mode 2 (blue circles) in (b) vs. $P$, when $\delta=0.8$ and $C=1$. The zoomed region in (b) is shown in (c). An example of time evolution of the field $|A|^2$ is shown in panel (d); the total energy (black curve), and modal energies $|C_{0,1,2}|^2$ (red, dark green, blue curves) are shown in panel (e), where $P=4.7$. 
  }
	\label{fig_chaoticons}
\end{figure}

In previous sections, we have discussed the emergence and features of DKS and breathers of different orders. 
However, another type of dynamical state exists, which is dramatically different from the former two. In prior work, we named such types of states as \textit{chaoticons} \cite{Sun2022b}, by following Ref.~\cite{verschueren_chaoticon_2014}. In this section, we perform a more detailed analysis of such states, and confirm their chaotic nature by means of the Lyapunov exponent method \cite{Rosenstein1993}.

In order to better address the study of these states, we show in Fig.~\ref{fig_chaoticons}(a) a bifurcation diagram, where the modification of the field energy with $P$ is illustrated for $\delta=0.8$. 
For small $P$, the systems exhibit stable DKS, which exist below the Hopf bifurcation point H$_{1,1}$. After crossing this point, symmetric breathers arise and persist in the interval $2.6<P<3.5$, until reaching H$_{1,2}$. Note that these bifurcations correspond to those in Figs.~\ref{fig_bif_pump_2p5}(a) and \ref{fig_bif_3p5}(a). Here we are just considering a different slice of the $(\delta,P)$-parameter space (see Section~\ref{sec_pha}).

At H$_{1,2}$, DKS stabilize again, and remain stable until $\rm C_2$ ($P\approx 4.6$). After this point, the total field energy exhibits significant random fluctuations [see Fig.~\ref{fig_chaoticons}], in contrast to DKS and breathers. These random fluctuations are characteristics of spatiotemporal chaotic behavior. Indeed, chaoticons exist in that region. An example of such a state is shown in Fig.~\ref{fig_chaoticons}(d), where we plot the time evolution of the field power $|A|^2$ for $P=4.7$. The chaoticon exhibits an irregular evolution, which is very different from that of breathers, as shown in Fig.~\ref{fig_breathers}.

This irregular behavior can also be observed by plotting the time evolution of the total and the modal energies, $E$ and $|C_n(t)|^2$, respectively [see Fig.~\ref{fig_chaoticons}(e)]. 
The formation of this dynamical state results from the localization effect that the potential has on the spatiotemporal chaotic dynamics, which is observed when $C=0$ \cite{Godey2014,parra-rivas_dynamics_2014}.
Note that in Fig.~\ref{fig_chaoticons}(a) chaoticons and stable DKS coexist in the interval $4.5\le P\le4.6$.

The Lyapunov exponents, which are obtained by linearizing the dynamics of finite-dimensional systems (in ODEs) around a given trajectory, can be used to measure the growth rate of generic small perturbations around such a trajectory, and therefore determine its chaotic nature \cite{manneville_dissipative_1995,clerc_quasiperiodicity_2013}. Although our system is infinite-dimensional (in PDE), its modal structure allows us to apply the previous approach in order to characterize the states that we have found. 
To do this, we need to run a sufficiently long simulation, until all solutions stabilize. In this way, we may obtain the time evolution of the mode energies [for examples in Fig.~\ref{fig_breathers}(i,2) and Fig.~\ref{fig_chaoticons}(e)]: from these, we may compute their Lyapunov exponents.

Figure~\ref{fig_chaoticons}(b) shows the modification of the largest Lyapunov exponent associated with modal energies $|C_0(t)|^2$ (i.e., $\lambda_0$, see {\color{red}$\circ$}) and $|C_2(t)|^2$ (i.e., $\lambda_2$, {\color{blue}$\circ$}) along the bifurcation diagram of Fig.~\ref{fig_chaoticons}(a). The theory states that steady-state attractors, such as solitons,
are characterized by negative exponents. Whereas complex behaviors, such as our chaoticons, exhibit positive exponents; time-periodic states (e.g., breathers) have zero Lyapunov exponents  \cite{clerc_quasiperiodicity_2013}. According to this theory, Fig.~\ref{fig_chaoticons}(b) clearly shows the existence of three different regimes (I-III), which we describe below:  
\begin{itemize}
\item[I]: DKS exist below H$_{1,1}$, as confirmed by a negative exponent;
\item[II]: breathers (between H$_{1,1}$ and H$_{1,2}$) have almost zero exponents ($<0.03$);
\item[III]: chaoticons exists for $P>4.5$, as confirmed their large positive exponents ($>45$).
\end{itemize}
Note that these computations are in perfect agreement with the bifurcation analysis that is shown in Fig.~\ref{fig_chaoticons}(a).



\begin{figure*}[tbp]
	\centering	\includegraphics[width=\textwidth]{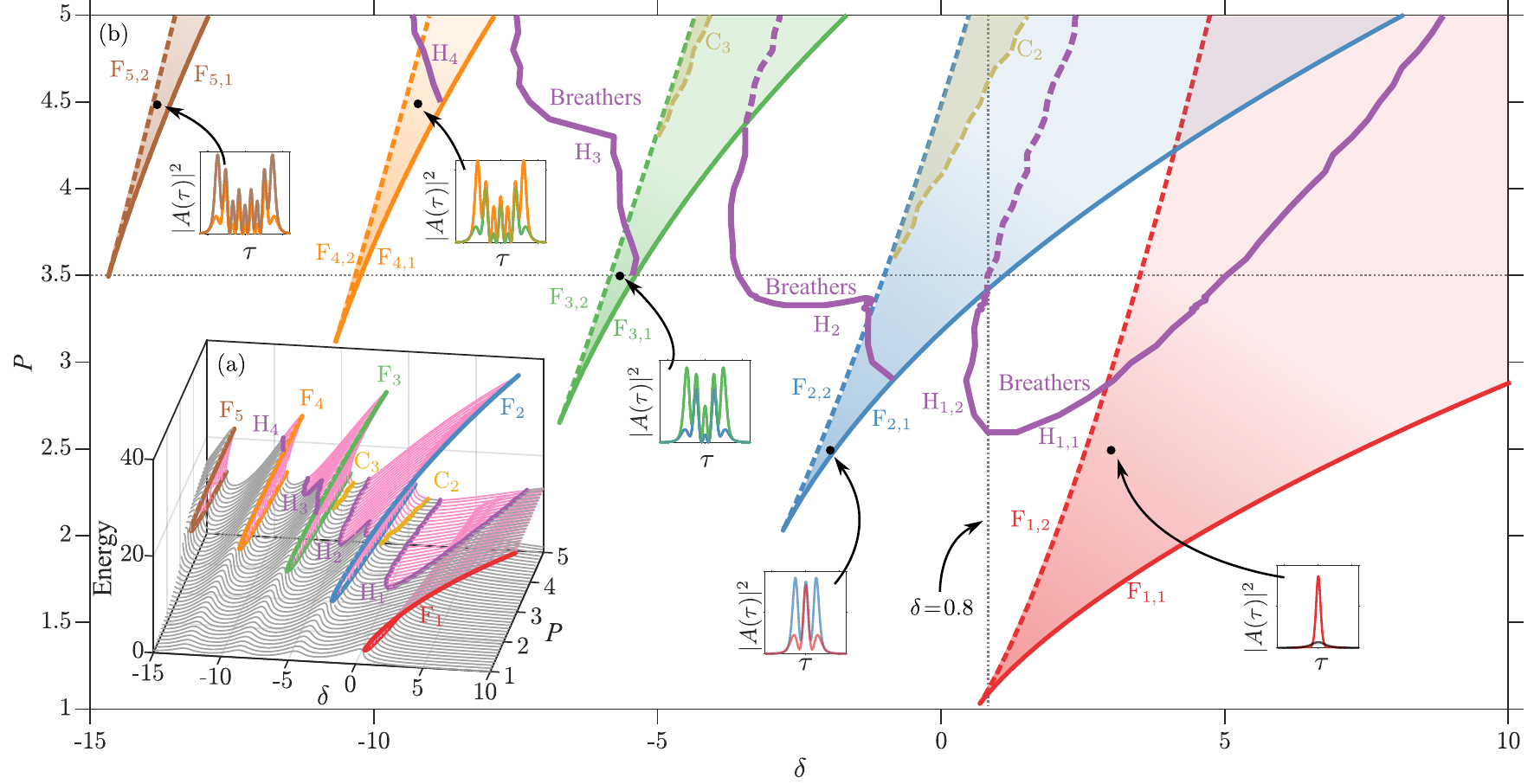}	
	\caption{
		Bifurcation diagrams of field energy vs. detuning $\delta$ for different pump strength $P$ in (a). Gray (pink) regions represent the stable soliton solutions of different orders (unstable solutions: breathers, chaoticons). The fold-points-connected curves ($\mathrm{F}_{m,1}$ and $\mathrm{F}_{m,2}$), the Hopf-points-connected curves ($\mathrm{H}_{m}$), and the curves of chaoticon onset ($\mathrm{C}_{m}$) at each resonance peak $m$ in (a) are represented in the phase diagram in (b), which is the top view of (a). 
		The solid (dashed) curve represents that the curve is at the top (lower) layer.
		On the one hand, increasing pumping folds the phase diagram, leading to the formation of bistable solutions (including solitons, breathers, and chaoticons). On the other hand, varying the detuning changes the solution order at different resonances surrounding each peak.
	}
	\label{fig_phase_diagram}
\end{figure*}

\section{Influence of pump: $\delta$ vs. $P$ phase diagram}\label{sec_pha}

So far, we have discussed the bifurcation structure of the solutions of Eq.~(\ref{eq_main}) by slicing the parameter space in two different ways. First, in Sections~\ref{DKS} and \ref{sec_bif_breathers} we have fixed the pump to the values $P=2.5,\,3.5$, and we studied the modification of the dynamics when varying $\delta$. Later, in Section~\ref{sec_chaoticon} we have fixed the detuning ($\delta=0.8$), and analyzed the modification of the stability of DKS when varying the pump strength. 

In this section, we unveil the organization of the different dynamical states of the system in the $(\delta,P)$-parameter space for $C=1$. Figure~\ref{fig_phase_diagram} summarize all our findings.

In Fig.~\ref{fig_phase_diagram}(a) we show, in a 3D representation, how the $E$ vs. $\delta$ bifurcation diagrams modify when the pump strength $P$ changes.
For $P<1.03$, there are four resonance peaks with low amplitude along $\delta$, which are located around the linear eigenvalues $\delta_n$ of the system. By increasing the strength of the pump, the first pair of fold points $\rm F_{1,1}$  and $\rm F_{1,2}$ is generated at the first resonance peak. These are represented by a red curve. 
By further increasing $P$, at higher resonance peaks, new fold bifurcation pairs nucleate (e.g., $\mathrm{F}_{2,m}$, $\mathrm{F}_{3,m}$, $\mathrm{F}_{4,m}$, with $m=1,2$). We marked these folds by using different colors. The pink regions show the unstable regimes, and are limited by different Hopf bifurcations (see purple curves).


The parameter space can also be characterized by projecting Fig.~\ref{fig_phase_diagram}(a) into the $(\delta,P)$-plane. This leads to the $(\delta,P)$-phase diagram which is shown in Fig.~\ref{fig_phase_diagram}(b), where we can easily differentiate several regions of different dynamical behavior. By inspecting this diagram, we may notice that the appearance of folds (therefore, of DKS bistable regions) at high-order resonance peaks requires larger values of the pump $P$, and that the bistable regions that they enclose widen with increasing $P$; so does the DKS energy.

Within each bistable region, two DKS of different order coexists for the same set of parameters, as illustrated in the five insets in Fig.~\ref{fig_phase_diagram}(b). In each case, higher-order DKS on each upper branch of the bifurcation diagram have two significant side peaks, when compared with the corresponding lower-order DKS located on the bottom branch 

With increasing $P$, the high-order DKS (on the top branch of each resonance peak) become unstable. This leads to the formation of breathers in the unstable regions, which are separated by the Hopf bifurcation curves (see purple lines).
Examples of this type of breathers are illustrated in Fig.~\ref{fig_breathers}(i,ii). The type of breathers which are located in the region surrounded by $\rm H_{2}$, on the second resonance peak, is shown in Fig.~\ref{fig_breathers}(iii-vii).
Similarly, higher-order breathers also appear at the top branches of the third and fourth resonance peaks, respectively, and they are marked by the curves $\rm H_3$, and $\rm H_4$ [see Fig.~\ref{fig_phase_diagram}(b)].

Chaoticons are located inside the region which is separated by the curves $\rm F_{2,2}$ and $\rm C_2$ (yellow dashed) on the second resonance peak, and the region between the curves $\rm F_{3,2}$ and $\rm C_3$ on the third resonance peak.
One can expect that for a larger pump, also this state will appear in s region of higher-order resonances.


\section{Influence of potential strength: $\delta$ vs. $\sqrt{C}$ phase diagram}\label{sec_pot_str}

\begin{figure}[t]
	\centering
	\includegraphics[scale=1]{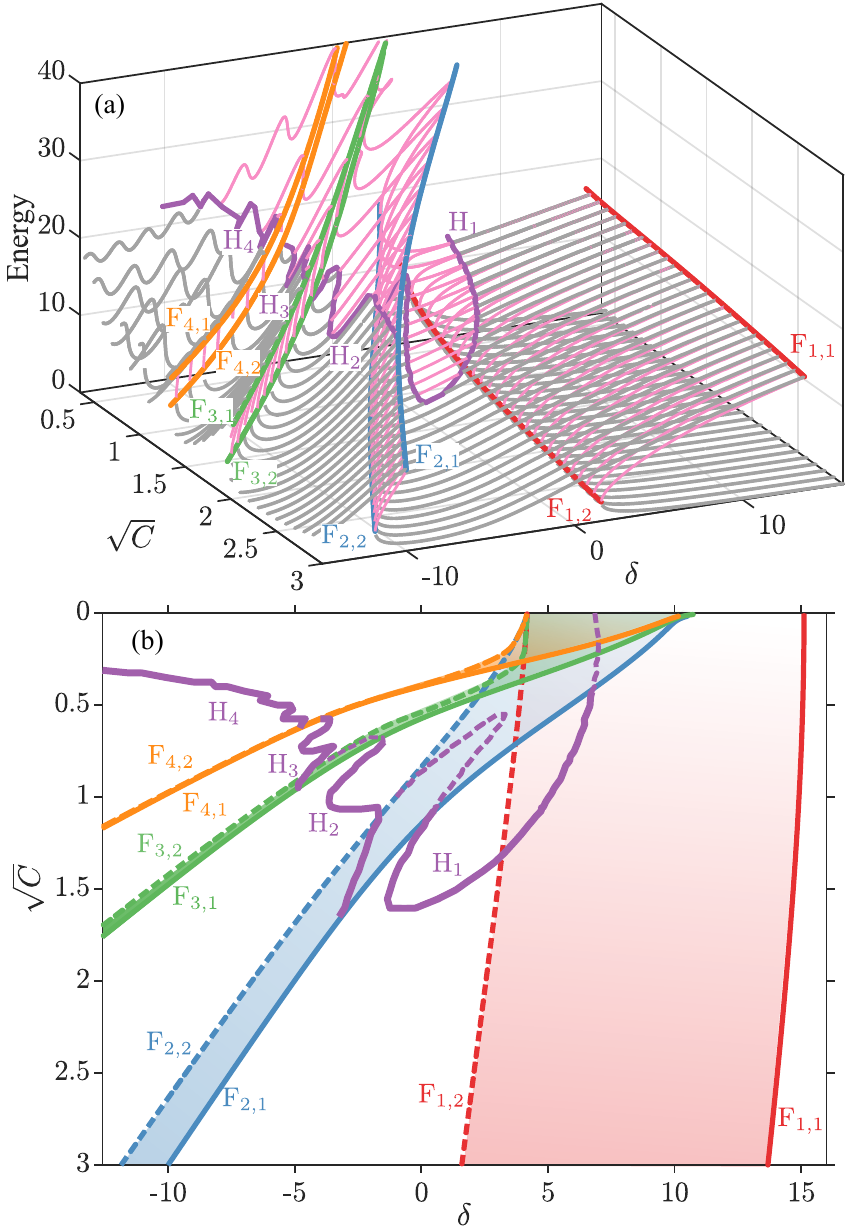}
		\caption{(a) Bifurcation diagrams of field energy vs. detuning $\delta$ for different potential strength $\sqrt{C}$ in (a), when $P=3.5$. (b) Phase diagram vs. detuning $\delta$ and potential strength $\sqrt{C}$, which is the top view of (a). The notations are the same as Fig.~\ref{fig_phase_diagram}. The importance of the potential in suppressing oscillatory instabilities and stabilizing DKS becomes apparent as $\sqrt{C}$ increases. As $\sqrt{C}$ decreases, the high-order resonance peaks converge to the same location, with the exception of the fundamental peak.
  }
		\label{fig_pha_dia_C}
\end{figure}

The parabolic potential strength $C$ [see Eq.~(\ref{eq_main})] has a very important impact on the system dynamics, since $C$ not only modifies the system eigenvalues, but it also varies the temporal width of the different states. Here, we discuss how the potential strength modifies the dynamics of the system, and we illustrate our findings by means of a $(\delta,\sqrt{C})$-phase diagram for fixed $P$. Here we have chosen $\sqrt{C}$ and not $C$, since the eigenvalues have a linear dependence on $\sqrt{C}$ [see Eq.~(\ref{eq_eigenvalue})].

Figure~\ref{fig_phase_diagram}(a) shows the modification of the $E$ vs. $\delta$ bifurcation diagrams when changing $\sqrt{C}$, for $P=3.5$.
Figure~\ref{fig_pha_dia_C}(b), in contrast, shows the projection of Fig.~\ref{fig_pha_dia_C}(a) onto the $(\delta,\sqrt{C})$-plane. As we can see, the positions of resonance peaks linearly increase with $\sqrt{C}$, which is consistent with our theoretical expectations [see Eq.~(\ref{eq_eigenvalue})]. However, the size of bistable regions (see fold bifurcations) remains almost fixed, since the tilting largely depends on the field peak power. Moreover, with increasing $\sqrt{C}$, the breathing regions gradually shrink, and eventually, they disappear. This indicates the relevance that the potential has in suppressing oscillatory instabilities and stabilizing DKS. 

From another perspective, reducing $\sqrt{C}$ impacts the systems in different ways, including the widening of the DKS unstable regions, the positive tilting of high-order resonances, and the broadening of the DKS states. 
Furthermore, by decreasing $C$, high-order resonance peaks converge to the same loci, except for the fundamental one. This transition may link to the homoclinic snaking type of structure which is observed for $C=0$ \cite{gomila_bifurcation_2007,parra-rivas_origin_2021}, although the confirmation of this conjecture needs further investigations.

\section{Conclusions}\label{sec_conclu}

In this work, we delve into the complex dynamics of dissipative wave structures in coherently-driven nonlinear Kerr cavities with a parabolic potential. The dynamics are governed by a modified Lugiato-Lefever equation model. We find that the potential significantly alters the system behavior. To show this, we have compared the stable states of the Kerr cavities in the absence and in the presence of the parabolic potential, by scanning the cavity detuning. 

Our results have revealed that the potential may stabilize complex spatiotemporal wave dynamics in favor of static DKS. Furthermore, we have discovered that the potential is also behind the emergence of DKS of higher orders.
To understand the formation of these higher-order DKS, we have first analyzed the linear eigenmodes of the system, and found that odd modes are suppressed owing to the coherent pumping. Whereas even modes are favored, exhibiting different excitation efficiencies. 

The inclusion of the  nonlinear Kerr effect tilts the resonance of these eigenmodes (towards a positive detuning), leading to bistable DKS of different orders. By projecting DKS on the eigenmodes, we demonstrated that the mode components of high-order DKS are primarily supported by the mode energy of corresponding orders.

Increasing the pump strength unveils the presence of breathers of different types and orders.  
Among them, asymmetric breathers emerge, as a result of the interaction of odd modes. 
In addition, we have discovered the existence of chaoticons (i.e., localized spatiotemporal chaotic  states), which have been characterized in terms of Lyapunov exponents.  

To summarize all of these complex dynamics, we have computed several phase diagrams in the $(\delta,P)$ and $(\delta,\sqrt{C})$-parameter space, which permits to organize and classify the different regimes and states of the dissipative wave structures. 


\section*{Acknowledgements}
	This work was supported by European Research Council (740355), Marie Sklodowska-Curie Actions (101064614,101023717), Sapienza University of Rome Additional Activity for MSCA (EFFILOCKER), Ministero dell’Istruzione, dell’Università e della Ricerca (R18SPB8227).


\appendix

\section{Spatial dynamical system for path-numerical continuation}\label{AppB}

This appendix shows how the path-continuation method is used to compute the bifurcations diagrams of this work by using AUTO-07p \cite{Doedel2009}.



Inserting 
\begin{equation} 
    A(\tau) = u(\tau) + \mathrm{i}v(\tau).
    \label{eq_re_im}
\end{equation}
into Eq.~(\ref{eq_main}) we obtain the coupled equations 
\begin{equation} 
\begin{split}
\partial_t u &= -\partial_\tau^2v -u-(u^2+v^2)v+\delta v +C\tau^2v+P,\\
\partial_t v &= +\partial_\tau^2u -v+(u^2+v^2)u-\delta u-C\tau^2u.
\end{split}
\label{eq_u_v}
\end{equation}
For steady-state solutions, these equations satisfy $\partial_tu=\partial_tv=0$; therefore, they are are solutions of the ordinary differential equations 
\begin{equation} 
\begin{split}
 -\dd_\tau^2v -u-(u^2+v^2)v+\delta v +C\tau^2v+P=0,\\
\dd_\tau^2u -v+(u^2+v^2)u-\delta u-C\tau^2u=0,
\end{split}
\label{eq_u_v2}
\end{equation}
where $\dd_{\tau}^2\equiv\frac{\dd^2}{\dd\tau^2}. $
By defining the new variables $u_1=u$, $u_2=v$, $u_3=\dd_\tau u=\dd_\tau u_1$ and $u_4=\dd_\tau v=\dd_\tau u_2$, $u_5=\tau$, the system of equations (\ref{eq_u_v2}) can be recast as the dynamical system:
\begin{equation} 
\begin{split}
\dd_\tau u_1&=u_3\\
\dd_\tau u_2&=u_4\\
\dd_\tau u_3&= -u_1-(u_1^2+u_2^2)v+\delta u_2 +Cu_5^2u_2+P\\
\dd_\tau u_4&
= -u_2+(u_1^2+u_2^2)u-\delta u_1-Cu_5^2u_1\\
\dd_\tau u_5&
= 1\\
\end{split}
\label{eq_auto}
\end{equation}
Due to the symmetry of the steady-state solutions, we can consider Neumann boundary conditions on just the half of domain, and write
\begin{equation} 
\begin{split}
u_3(0)&=\dd_\tau u(0)=0,\\
u_3(R)&=\dd_\tau u(R)=0,\\
u_4(0)&=\dd_\tau v(0)=0,\\
u_4(R)&=\dd_\tau v(R)=0,\\
u_5(0)&=0,\\
\end{split}
\label{eq_auto_b}
\end{equation}
where $R$ is the half of domain size. The boundary value problem defined by  Eq.~(\ref{eq_auto}) and the boundary conditions (\ref{eq_auto_b}) can then be solved using AUTO-07p, as described in Ref.~\cite{doedel_numerical_1991}.

\section{Ordinary differential equations of the modal}\label{App_A}

This appendix shows the derivation of the set of modal differential equations [Eq.~(\ref{eq_couple_mode})] from the partial differential equation [Eq.~(\ref{eq_main})].

First, we simplify some terms in Eq.~(\ref{eq_main}).
To do so, substituting Eq.~(\ref{A_ex}) into the left hand side of Eq.~(\ref{eq_main}) yields 
\begin{equation} 
	\begin{split}
	\frac{\partial A}{\partial t}& = \frac{\partial }{\partial t}\left[\sum_{n=0}^NC_n(t)e^{i\delta_nt}{ \psi}_n(\tau)\right]\\
	& =\sum_{n=0}^N \frac{\dd C_n(t)}{\dd t}e^{i\delta_nt}{\psi}_n(\tau)  
	+\sum_{n=0}^NC_n(t)e^{i\delta_nt} \cdot i\delta_n{\psi}_n(\tau).
	\end{split}
\label{eq_ap_1}
\end{equation}
By substituting Eq.~(\ref{A_ex}) into the first two terms on the right-hand side of Eq.~(\ref{eq_main}), and using Eq.~(\ref{eq_eigen_fu_va}), we have 
\begin{equation} 
	\begin{split}
		\hat{H_0}A &= \hat{H_0}\sum_{n=0}^NC_n(t)e^{i\delta_nt}{ \psi}_n(\tau)\\
		 &= \sum_{n=0}^NC_n(t)e^{i\delta_nt}\cdot\hat{H_0}{ \psi}_n(\tau)\\ 
		 & =\sum_{n=0}^NC_n(t)e^{i\delta_nt}\cdot i\delta_n{\psi}_n(\tau).
	\end{split}
\label{eq_ap_2}
\end{equation}
Then, we substitute Eq.~(\ref{eq_ap_1}) and Eq.~(\ref{eq_ap_2}) into Eq.~(\ref{eq_main}),
and 
implementing algebraic simplification, we obtain
\begin{equation} 
	\begin{split}
		\sum_{n=0}^{\infty}\frac{\dd C_n(t)}{\dd t}e^{i\delta_nt}{\psi}_n(\tau) = \hat{H}_1A+P,
	\end{split}
	\label{eq_03}
\end{equation}
where $\hat{H}_1= \mathrm{i}|A|^2  - (1+\mathrm{i}\delta)$. Next, we project Eq.~(\ref{eq_03}) on mode $n$, 
\begin{equation} 
	\begin{split}
		\int_{-\infty}^{\infty}\sum_{m=0}^{\infty}\frac{\dd C_m(t)}{\dd t}e^{i\delta_mt}{\psi}_m(\tau)&{\psi}_n(\tau)\dd \tau =\int_{-\infty}^{\infty}\left[ \hat{H}_1A+P\right] {\psi}_n(\tau)\dd \tau.
	\end{split}
\end{equation}
Finally, by doing the simplification, we end up with the set of coupled mode ODEs of the model
\begin{equation} 
\begin{split}
    \frac{\dd C_n(t)}{\dd t} =	&e^{-i\delta_nt}\int_{-\infty}^{\infty}P\psi_n(\tau)\dd \tau\\
    +&e^{-i\delta_nt}\int_{-\infty}^{\infty}\mathrm{i}|A|^2A\psi_n(\tau)\dd \tau\\
    -&e^{-i\delta_nt}(1+\mathrm{i}\delta)C_n
\end{split}
\end{equation}
where each ODE governs the dynamics of one mode. 

\bibliographystyle{apsrev4-2}

\bibliography{main}





\end{document}